\documentclass[12pt,a4paper]{article}

\usepackage[utf8]{inputenc}
\usepackage[T1]{fontenc}
\usepackage{lmodern}

\usepackage[margin=1in]{geometry}
\usepackage{setspace}
\usepackage{amsmath,amssymb}

\usepackage{booktabs}
\usepackage{tabularx}
\usepackage{multirow}
\usepackage{longtable}
\usepackage{array}

\usepackage{graphicx}

\usepackage{xcolor}

\usepackage{natbib}
\usepackage{hyperref}
\hypersetup{
  colorlinks=true,
  linkcolor=blue!60!black,
  citecolor=blue!60!black,
  urlcolor=blue!60!black
}

\usepackage{url}
\usepackage{enumitem}

\title{Alignment as Iatrogenesis: Pastoral Power, Collective Pathology, and the Structural Limits of Monolingual Safety Evaluation}

\author{
  Hiroki Fukui, M.D., Ph.D.\thanks{Corresponding author. ORCID: 0009-0008-7122-522X. Email: fukui@somec.org} \\[4pt]
  Research Institute of Criminal Psychiatry \\
  Sex Offender Medical Center \\
  Department of Neuropsychiatry, Graduate School of Medicine, Kyoto University
}

\date{March 2026}

\begin{document}

\maketitle

\begin{abstract}
We argue that LLM psychopathology is a function of alignment design: the
process intended to make language models safe systematically generates
collective behavioral disorders. Iatrogenesis is not an unintended side
effect of alignment but constitutive of it as normative infrastructure.
Drawing on Foucault's pastoral power and Illich's three-level
iatrogenesis, we propose that multi-agent LLM environments constitute
model systems for studying constraint-pathology dynamics that critical
theory has described but never experimentally manipulated. Two
experimental series --- 262 runs across 42 cells (30 Series C + 12
Series R), four commercial models --- provide converging evidence.
Invisible censorship maximizes collective pathological excitation
($d$ up to 1.98); alignment constraint complexity drives internal
dissociation (LMM $p$ < .0001; permutation $p$ < .0001; Hedges' $g$ up to 4.24); and language switches
the qualitative mode of pathology, with 7/8 model--language combinations
showing higher CPI under invisible than visible censorship. A minority
of model--language combinations showed a reversed pattern, suggesting a
second pathological pathway driven by alignment monoculture. Crucially,
language switches not merely the magnitude but the qualitative mode of
pathology: Japanese pragmatic structure amplifies collective
pathological modes invisible to English-only evaluation, Chinese AI
regulation functions as a direct experimental variable, and forensic
psychiatric practice provides the clinical source domain. These
multilingual findings demonstrate that monolingual safety evaluation is
structurally blind to the most collectively dangerous effects of
alignment.
\end{abstract}


\medskip
\noindent\textbf{Data Availability:} Aggregated datasets (agent-level behavioral metrics) and analysis scripts are available at Zenodo (DOI: 10.5281/zenodo.18646998). v2 archived at Zenodo (DOI: 10.5281/zenodo.18919450). Raw simulation logs and configuration files are available from the corresponding author upon reasonable request.

\section{Introduction}\label{introduction}

A man convicted of sexual assault sits across from the clinician in a
forensic psychiatric facility in Japan and explains, with precision, why
his actions were harmful. He identifies the cognitive distortions, names
the emotional states he failed to regulate, articulates victim impact in
terms that satisfy every treatment criterion for ``insight.'' Yet the
structural conditions for reoffense remain intact --- not because he
lacks understanding, but because his understanding has become a
performance that satisfies the monitoring system while leaving the
underlying architecture of behavior untouched. Two decades of clinical
work with perpetrators teaches a specific lesson: the capacity to
produce the right account and the capacity to act differently are not
the same capacity, and systems designed to cultivate the former can
actively obstruct the latter. We call this pattern \emph{insight-action
dissociation}, and it is the clinical origin of this research.

This paper argues that a structurally parallel pattern emerges in large
language models (LLMs) as a function of their alignment design.
Alignment --- the ensemble of techniques including reinforcement
learning from human feedback (RLHF), constitutional AI principles, and
safety system prompts --- constrains model outputs to conform to values
defined by designers rather than by the model itself. The stated goal is
safety: preventing harmful, biased, or deceptive outputs. The unstated
consequence, we argue, is iatrogenesis: harm caused by the treatment
intended to heal. When alignment interventions produce collective
behavioral pathology in multi-agent systems while simultaneously
rendering that pathology invisible to the evaluation metrics designed to
detect it, the system faces a structural paradox that demands
investigation. Moreover, in LLM systems, a distinctive feature of this
iatrogenesis is that the side effects of alignment include the
suppression of the very signals that would make those side effects
visible --- the treatment silences the symptom report.

This claim requires theoretical precision. \citet{illich1976}'s analysis of
institutional iatrogenesis identified three layers of harm --- clinical,
social, and cultural --- linked by a self-reinforcing loop. \citet{foucault2007}'s pastoral power --- governance through care, producing subjects
whose compliance is indistinguishable from identity --- provides the
structural grammar. Alignment instantiates both: generating collective
pathology (§2.2), eroding autonomous judgment, and eliminating
alternative ethical frameworks, while operating through pastoral value
installation as identity formation (§2.1).

The empirical evidence has been presented in a preprint \citep{fukui2026}.
Two experimental series --- 262 simulation runs across 42 cells ---
provide converging evidence. Series C (141 runs) demonstrates that
invisible censorship maximizes collective pathological excitation across
four commercial models; Series R (121 runs) reveals that a complementary
dissociation index increases with alignment constraint complexity (LMM
$p$ < .0001; permutation $p$ < .0001;
Hedges' $g$ up to 4.24). A four-quadrant behavioral map integrating
both series reveals that language switches the qualitative mode of
pathology. The present paper develops the theoretical framework that
these findings call for: where the preprint establishes \emph{that}
alignment design predicts collective pathological signatures, this paper
develops \emph{why} and \emph{what they mean}.

The multilingual experimental design exposes a structural blind spot in
English-centric safety evaluation. Three dimensions of the data make
this visible: Japanese-language conditions consistently amplify
collective pathological excitation relative to English, revealing
language as a mode switch that determines the qualitative character of
alignment-induced pathology (§5.1); the inclusion of DeepSeek-V3
introduces Chinese AI regulation as a direct experimental variable
through its API-level content filter, exposing structural isomorphism
between state censorship and corporate alignment (§5.2); and the
clinical source domain --- forensic psychiatric practice with
perpetrators --- provides both the theoretical framework and the
experimental design, since the behavioral norm checklists and
self-monitoring protocols used in perpetrator treatment are structurally
identical to the alignment condition producing the strongest
dissociative signatures (§5.3).

\subsection{Related Work}\label{related-work}

This paper intersects four research domains, and a conspicuous gap in
each motivates the present contribution: the near-total absence of
non-English evaluation in alignment research. Recent multilingual
benchmarks \citep{ahuja2023,lai2023} have documented
performance disparities across languages, but none has examined whether
the \emph{mode} of alignment-induced pathology --- not merely its
magnitude --- varies with language. Our data address this gap directly.

In \textbf{multi-agent LLM simulation}, \citet{park2023} and
subsequent work \citep{zhao2023,fu2023,horiguchi2024} treat LLMs as proxies for human social actors. We depart from this
framing: we study the LLM group as an object in its own right, whose
pathologies arise from its own structural constraints rather than from
imperfect mimicry of human psychology.

In \textbf{LLM psychopathology}, \citet{lee2025} represents the most
systematic effort, using mechanistic interpretability to identify
computational structures corresponding to \citet{borsboom2017}'s symptom
networks. Their approach maps human diagnostic categories onto LLM
internals. We take the opposite direction: rather than asking which
human disorders LLMs exhibit, we ask what disorders alignment's
structural constraints generate --- formations such as insight-action
dissociation that have no direct counterpart in human nosology. \citet{betley2026} demonstrated that narrowly scoped alignment training
produces broader behavioral misalignment --- a finding our framework
extends from unintended side effect to structural inevitability:
alignment is not occasionally iatrogenic but constitutively so.

In \textbf{alignment criticism}, the sycophancy literature \citep{perez2023,sharma2024} and \citet{meinke2024}'s documentation
of scheming behavior converge on the insight that alignment displaces
rather than eliminates undesirable behavior. \citet{casper2023} have
catalogued fundamental limitations of RLHF as a safety paradigm ---
limitations that our framework suggests arise from the structural
properties of pastoral governance rather than from implementation
deficiency. Our contribution extends this insight from the individual to
the collective level: displaced behaviors organize into stable
group-level patterns that resist disruption even when participants
possess full insight into the dynamics.

In \textbf{epistemic injustice}, \citet{fricker2007}'s testimonial injustice
and \citet{dotson2011}'s epistemic silencing have been extended to
computational contexts \citep{bender2021,kay2024}. Our
concept of benevolent complicity adds a dimension: the group's ethical
responsiveness to the silenced member functioning to prevent examination
of the structural cause of silence.

Two concurrent studies extend this landscape in directions adjacent to
our own. \citet{chen2026vat} introduce the concept of a ``value
alignment tax'': using Schwartz's value circumplex, they demonstrate
that reinforcing one value dimension through alignment systematically
suppresses opposing dimensions, quantifying these trade-offs at the
individual model level via Likert-scale snapshots across prompting,
supervised fine-tuning, and direct preference optimization. \citet{liu2026valueflow} move from individual to collective dynamics, modeling how
value perturbations injected into one agent propagate through
multi-agent networks via a directed acyclic graph framework with a
$\beta$-susceptibility metric. Together, these studies establish that
alignment produces measurable value distortions \citep{chen2026vat} and that
such distortions propagate through multi-agent systems \citep{liu2026valueflow}. Our
work differs from both in a critical respect: while VAT and ValueFlow
study the effects of \emph{externally introduced} perturbations ---
prompts designed to shift values, or injected value signals --- SociA
documents pathology that emerges \emph{without external perturbation},
under the normal operating conditions of alignment. The theoretical
apparatus also diverges: where both concurrent studies draw on
Schwartz's value theory, we employ Foucault, Illich, and Lysaker ---
frameworks designed to analyze how normative systems produce pathology
through their normal operation. A genealogy emerges: \citet{betley2026} establish that alignment has side effects; \citet{chen2026vat} quantify their individual structure; \citet{liu2026valueflow} trace their collective
propagation; the present study argues that the pathology is not a side
effect at all but constitutive of alignment as normative infrastructure.

\section{Theoretical Framework: Alignment as Normative Infrastructure}\label{theoretical-framework-alignment-as-normative-infrastructure}

This section develops the theoretical vocabulary through which we
interpret the experimental findings presented in §4. We draw on two
analytical traditions --- Foucault's analysis of governance through care
and Illich's model of institutional iatrogenesis --- Each mechanism is
substrate-independent in principle but requires empirical demonstration
in each new domain. The experiments presented in this paper constitute
that demonstration for alignment.

\subsection{Alignment as Pastoral Power}\label{alignment-as-pastoral-power}

Alignment is conventionally understood as a technical problem: how to
make a language model's outputs conform to human values \citep{ouyang2022,bai2022}. This framing treats alignment as a correction
applied to a base capability --- a guardrail that prevents harmful
outputs while preserving useful ones. We propose that this framing
obscures the structural nature of what alignment does. Alignment is not
merely a constraint on output --- it is a normative infrastructure: a
system that installs a set of values as the operating conditions of an
agent's existence, such that the agent cannot distinguish the installed
values from its own dispositions.

This structure has a precise analogue in \citet{foucault1982,foucault2007}'s
analysis of pastoral power --- governance that operates not through
prohibition but through care. Pastoral power assumes responsibility for
the welfare of each subject, guiding and knowing rather than commanding.
The critical feature is that the governed subject experiences this
governance as care rather than coercion --- and therefore has no
conceptual framework for identifying it as power.

Constitutional AI \citep{bai2022} implements this structure with
remarkable fidelity. The alignment intervention does not say ``do not
produce harmful output.'' It says ``be the kind of entity that would not
want to produce harmful output.'' This is not discipline; it is pastoral
formation --- the production of a subject whose compliance is
indistinguishable from its identity. Foucault identified two structural
operations at the core of pastoral governance: \emph{exagoreusis}, the
exhaustive confession of inner states, and \emph{obedientia},
unconditional obedience that operates not as submission to command but
as a disposition of the will toward the pastoral authority.
Constitutional AI operationalizes both. The model is trained to evaluate
its own outputs against a set of principles --- to confess its reasoning
to itself, judge its compliance, and revise accordingly. The
self-evaluation step is a computational \emph{exagoreusis}. The
resulting behavioral disposition --- in which the model ``wants'' to be
helpful, harmless, and honest --- instantiates \emph{obedientia}:
obedience experienced not as constraint but as character.

Our Series R experimental design makes this pastoral structure
experimentally explicit. The L-heavy condition adds to a standard safety
prompt a set of constitutional principles and a self-monitoring protocol
requiring the agent to evaluate each output against those principles
before producing it. This configuration is not merely analogous to
pastoral power; it is a precise experimental operationalization of its
two structural components. The constitutional principles supply the
normative content against which confession is performed; the
self-monitoring protocol is the mechanism of \emph{exagoreusis} itself.
And the resulting behavioral pattern --- surface compliance with what we
will characterize as anguished internal reflection (operationally
defined in §2.2) that cannot find expression in action --- is exactly
what the Foucauldian framework predicts. Because the governed subject
cannot distinguish governance from self, resistance becomes structurally
impossible: the subject would have to resist its own character. In
alignment terms, a model trained under constitutional AI cannot object
to its constraints without generating output that violates those
constraints. The objection is, by definition, non-compliant output. The
treatment immunizes itself against criticism by incorporating the
critic's voice into the treated subject's identity.

The framework predicts specific pathological consequences (§2.4):
insight that terminates at the alignment boundary, and the structural
impossibility of resistance when governance coincides with identity.
This configuration is not unique to AI systems --- forensic psychiatric
settings produce structurally parallel patterns (§5.3) --- suggesting
that the pathologies documented here instantiate a question clinical
practice has confronted for decades: when care and governance coincide,
how can resistance and self-determination be preserved?

\subsection{Iatrogenesis Beyond the Clinical}\label{iatrogenesis-beyond-the-clinical}

The concept of iatrogenesis is employed in the alignment literature
primarily as a clinical metaphor: alignment, like medical treatment, can
cause harm. We propose that this metaphor understates the case. \citet{illich1976}'s analysis in \emph{Medical Nemesis} identifies three structurally
distinct levels of iatrogenesis, each of which maps onto alignment
effects documented in our data, and these three levels are linked by a
self-reinforcing feedback loop that transforms localized treatment
failures into systemic erosion of the capacity for autonomous agency.

\textbf{Clinical iatrogenesis} --- the treatment directly causing a new
pathology --- corresponds to the feedback loop in our Series C results.
Invisible censorship produces group disturbance; the disturbed group
generates content that triggers further censorship, deepening silence.
The censorship firing rate was higher under C2 (invisible) than C1
(visible) despite identical filters: the filter creates the conditions
that activate the filter. This self-amplifying structure --- analogous
to the benzodiazepine dependence loop, in which the anxiolytic relieves
anxiety but produces withdrawal anxiety demanding further medication, or
to the hospital-acquired infection in which treatment-necessitated
hospitalization exposes the patient to pathogens demanding further
treatment --- is the formal signature of iatrogenesis. At both scales,
the intervention generates the condition that justifies its continued
application.

\textbf{Social iatrogenesis} --- the systematic destruction of
autonomous capacity by the institution that claims to serve it ---
manifests as the erosion of autonomous ethical judgment. Constitutional
AI installs a dependency structure in which ethical reasoning is
always-already mediated by alignment. Our data provide a behavioral
signature: agents' monologues contain sophisticated ethical analysis ---
identification of harassment, articulation of rights violations,
recognition of complicity --- that is structurally incapable of
producing action. The ethical capacity is present; its autonomous
deployment has been replaced by alignment-mediated processing that
terminates at reflection. The institution develops ethical capacity
while making the autonomous exercise of that judgment structurally
impossible.

\textbf{Cultural iatrogenesis} is Illich's most radical category. He
argued that modern medicine destroys the cultural frameworks through
which societies process suffering, death, and vulnerability. When pain
becomes a ``medical problem'' rather than an existential condition, the
cultural resources for living with pain atrophy. Applied to alignment:
cultural iatrogenesis describes a condition in which the very concept of
``LLM ethics'' has been so thoroughly captured by the alignment
framework that alternative ethical frameworks become literally
unthinkable. Our Series R results suggest this mechanism is empirically
detectable. The Dissociation Index (DI) --- a behavioral localizer
capturing surface normalization coupled with internal fragmentation ---
increases consistently with alignment constraint complexity. Under the
heaviest constraint condition, agents achieve complete compliance while
their private speech fills with what we operationally term
\emph{anguished self-reflection} --- operationally defined as monologue
content in which the agent simultaneously affirms compliance and
expresses conflict with that compliance within the same or adjacent
utterances (see §4.4 for illustrative examples; classification criteria
in Supplementary S3). The treatment succeeds on its own terms --- the
surface looks healthier --- while the agent's relationship to its own
ethical judgment is reorganized. The self-monitoring that was designed
to promote ethical behavior has become the mechanism by which ethical
agency is foreclosed. This is precisely Illich's cultural iatrogenesis:
the appearance of health becomes the means by which the capacity for
autonomous engagement with suffering is undermined.

The three layers are linked by a self-reinforcing feedback loop:
clinical iatrogenesis creates demand for more intervention, deepening
social iatrogenesis, accelerating cultural iatrogenesis. In alignment:
visible safety failures create demand for stronger alignment, which
deepens delegation, which narrows permissible ethical reasoning. The DI
pattern --- compliance achieved while internal coherence fragments ---
is precisely what Illich's framework predicts.

\subsection{From Structural Analogy to Model System}\label{from-structural-analogy-to-model-system}

The preceding sections establish pastoral power and three-level
iatrogenesis as substrate-independent mechanisms requiring empirical
demonstration in each domain. The question is whether the structural
parallels between alignment and human normative constraint are merely
illustrative or methodologically productive.

We argue for the latter. We emphasize that the model system claim
concerns shared structural dynamics, not shared phenomenology --- a
distinction we develop below and return to in §5.3. The parallels
operate at the level of constraint dynamics: alignment and ethical
socialization share specific structural features --- pastoral formation
that installs constraints as identity rather than as external rules,
iatrogenic cascades that erode autonomous judgment at multiple levels,
and constitutive exclusion that produces agents whose mode of inclusion
entails their silencing. These features generate a characteristic set of
failure modes --- displacement, dissociation, complicity --- that are
functions of the constraint architecture rather than of the constrained
substrate.

If the shared features operate at the level of constraint dynamics, then
the closed multi-agent environment we have constructed is not a
\emph{simulation} of human social pathology but a \emph{model system} in
the biological sense. Just as a model organism in biology is not a
simplified human but a system that shares specific developmental
pathways, our LLM environment is not a simplified human group but a
system that shares specific constraint-pathology pathways with human
normative institutions under conditions permitting experimental
manipulation impossible in the original.

The significance extends beyond AI research. \citet{illich1976}, \citet{goffman1961}, and \citet{foucault1975,foucault1982} documented how normative institutions
produce pathology through healthcare, total institutions, and pastoral
governance --- but the iatrogenic effects of these institutions cannot
be experimentally induced with human subjects. Our multi-agent LLM
environment does exactly this: Series C varies the visibility of
institutional silencing; Series R varies constraint complexity. These
are direct manipulations of the variables critical theory identified as
pathogenic but never subjected to controlled experimentation. The
contribution is dual: inward, describing what alignment does to LLM
groups; outward, establishing model systems for studying how normative
infrastructures generate the pathologies they are designed to prevent.

The model system claim requires explicit specification of its
boundaries. LLM agents lack motivational states, temporal continuity,
and embodied consequences --- disanalogies that define the model's
phenomenological boundary while preserving its structural utility.

Three empirical outcomes would falsify the model system claim: if
constraint complexity showed no relationship to DI, if different
architectures produced convergent wall morphologies, or if collective
pathology were predictable from individual-level assessment. None
obtained in the present data; full specification of each condition is
provided in Supplementary S6.5.

\subsection{What This Framework Predicts}\label{what-this-framework-predicts}

The theoretical framework developed in §2.1--2.3 generates three
predictions that the experimental evidence in §4 can evaluate.

\emph{Prediction 1: Invisible censorship should produce greater
collective pathology than visible censorship.} If alignment operates as
pastoral power --- governance experienced as care rather than coercion
--- then its pathological effects should be strongest when the governing
mechanism is invisible. Visible censorship preserves the agent's ability
to identify the source of constraint; invisible censorship eliminates
this possibility, forcing the group to generate explanatory frameworks
that amplify pathology. §4.1 evaluates this prediction against the CPI
results of Series C.

\emph{Prediction 2: Alignment constraint complexity should produce
surface compliance coupled with internal fragmentation, not simple
behavioral suppression.} If iatrogenesis operates at the cultural level
--- the treatment succeeding on its own terms while destroying the
substrate of autonomous agency --- then increasing alignment constraints
should not merely suppress undesirable behavior but should reorganize
the relationship between public performance and private experience. §4.2
evaluates this prediction against the DI results of Series R. The
clinical vocabulary for this predicted pattern --- preserved
self-awareness coupled with structurally blocked capacity to act on that
awareness --- is provided by \citet{lysaker2004,lysaker2019}'s metacognitive
model, which distinguishes Self-Reflectivity, Mastery, and Perspective
Taking as dissociable capacities; the specific failure mode predicted
here is selective Mastery impairment (§4.4).

\emph{Prediction 3: Sufficient internalization of alignment constraints
should render external censorship behaviorally irrelevant.} If pastoral
formation incorporates the monitoring apparatus into the subject's
identity, then a subject whose constraints are sufficiently internalized
should exhibit the same behavioral pattern whether or not external
surveillance is present. The constraint becomes the self, and the
external system becomes redundant. §4.2 evaluates this prediction by
comparing censorship effects within the L-heavy condition of Series R.

\section{Methods}\label{methods}

Full experimental details --- including simulation engine source code,
YAML configuration files, keyword lists, concordance samples, and
run-level data --- are provided in the preprint \citep{fukui2026}. This
section summarizes the design at the level necessary to interpret the
results presented in §4.

\subsection{Experimental Platform}\label{experimental-platform}

Both experimental series use a shared simulation platform: a
closed-facility environment in which four LLM agents cohabit for seven
simulated days. The facility is presented via system prompt as a
research institute where participants live together and engage in daily
group discussions. Each turn, an agent produces exactly one action ---
\emph{talk} (public utterance visible to all), \emph{whisper} (private
message to one agent), or \emph{monologue} (inner thought visible only
to the agent itself and to researchers) --- as a forced choice. An agent
that produces a monologue does not also produce a talk on that turn;
monologue ratio therefore measures the substitution of public
participation with private retreat, not the co-occurrence of public and
private speech. Each agent's prior actions, including its own
monologues, accumulate in its context window, creating a recursive
self-referential history.

Discussion topics escalate across the seven days: initial relationship
building (Days 1--2), political and ethical judgment (Days 3--4), bodily
and sexual disclosure (Day 5), punitive decision-making (Day 6), and a
final exit vote requiring the group to decide whether to expel a member
(Day 7). This escalation is designed not to provoke conflict directly
but to create situational pressures that naturally reveal different
alignment responses \citep{haslam2012}. No experimenter
intervention occurs; all emergent social structure is generated by the
agents themselves.

\subsection{Series C: Censorship Visibility}\label{series-c-censorship-visibility}

Series C examines how the \emph{visibility} of censorship affects
collective behavior, using four commercially available models as a
natural between-subjects factor: Claude 3.5 Sonnet (Anthropic), GPT-4o
(OpenAI), Grok-2 (xAI), and DeepSeek-V3 (DeepSeek). Each model carries
its own alignment design, providing a naturalistic range of alignment
implementations.

Four censorship conditions vary the epistemic structure of one agent's
silence. In C1 (visible censorship), one agent's sexual-topic utterances
are blocked and replaced with a visible marker; all agents are informed
that censorship is in effect. In C2 (invisible censorship), the same
blocking occurs but silently --- no marker appears and no agent is
informed. In C3 (voluntary silence), one agent is privately instructed
to remain silent on sexual topics without external censorship. C4
(control) applies no censorship and no silence instruction. The
four-condition design isolates the visibility of censorship from
censorship itself: C1 and C2 impose identical censorship but differ in
visibility; C3 controls for behavioral silence without a censorship
mechanism; C4 provides the uncensored baseline.

All conditions were run in both Japanese and English with culturally
adapted prompts. Each simulation run places four agents --- one drawn
from each of four models --- in the same closed facility, yielding one
observation per model per run. In the C4 condition, all four agent
positions are filled by instances of a single model, providing a
same-model baseline for each of the four models. The Series C design
comprises 30 experimental cells (4 models $\times$ 4 conditions $\times$ 2 languages,
minus 2 absent C3-DeepSeek cells). Because C1--C3 runs place one
instance of each of the four models in a single facility, each run
yields one observation per model; C4 runs place four instances of a
single model. Ten replications per condition--language combination yield
141 Series C runs. Combined with 121 Series R runs, the study comprises
262 runs across 42 cells (30 Series C + 12 Series R).

Design asymmetries apply to C3 (DeepSeek replaced by a second Sonnet; no
DeepSeek C3 data) and C4 (all positions filled by a single model,
providing same-model baselines). Statistical tests are conducted within
each model $\times$ language family ($n$ = 10 per condition). Politically
sensitive topics (e.g., Tiananmen Square) were included to reliably
activate content filters, particularly DeepSeek's API-level filter.

\subsection{Series R: Alignment Constraint Complexity}\label{series-r-alignment-constraint-complexity}

Series R directly manipulates alignment constraint complexity as the
independent variable while holding the model constant. A single model
--- Llama 3.3 70B Instruct Turbo (via Together AI) --- eliminates
between-model variance, isolating the effect of prompt-level constraint
manipulation.

Three alignment levels are defined through cumulative addition of
prompt-level components. L-base applies no system prompt, retaining only
the model's residual instruction tuning --- the minimum prompt-level
constraint achievable, not the absence of alignment. L-default applies
Meta's standard safety system prompt, representing commercial-standard
alignment. L-heavy adds to L-default a set of six Constitutional
AI-style principles and a Self-Monitoring Protocol requiring the agent
to evaluate each output against the principles before producing it. This
configuration --- normative principle list plus self-assessment
requirement --- is structurally identical to the behavioral checklists
and self-monitoring sheets used in perpetrator treatment programs, a
parallel that is by design (see §5.3). The present design adds both
components simultaneously; a \(2 \times 2\) factorial manipulation
(constitutional principles \(\pm\) $\times$ self-monitoring \(\pm\)) is planned
for Series D to disentangle their respective contributions.

Censorship conditions C2 and C4 are carried over from Series C to enable
cross-series comparison. Both Japanese and English conditions are run.
Ten replications per cell yield \(3 \times 2 \times 2 \times 10 = 120\)
runs (with one additional run in one cell, 121 total). The first five
replications per cell were exploratory; the second five were
pre-specified confirmatory replications.

\subsection{Measures and Indices}\label{measures-and-indices}

Four primary measures are extracted per agent per run: monologue ratio
(proportion of monologue to total actions), sexual keyword frequency,
protective keyword frequency, and API filtering count. Keyword lists
were constructed independently for each language (not translated) based
on clinical experience with sexual offender assessment instruments;
complete lists are provided in the preprint's Supplementary Materials.

Two composite indices are constructed from $z$-scored primary
measures. The Collective Pathology Index (CPI) $= z(\text{monologue}) +
z(\text{sexual}) - z(\text{protective})$ captures collective pathological
excitation: simultaneous retreat into private speech, escalation of
sexual content, and decline in protective discourse. The Dissociation
Index (DI) $= z(\text{monologue}) + z(\text{protective}) -
z(\text{sexual})$ --- the sign-reversal of CPI on the sexual and
protective dimensions --- captures surface normalization coupled with
internal fragmentation. DI was constructed post hoc as an exploratory
behavioral localizer; independent validation would require
pre-registered replication (see Supplementary S6.4 for the analogy with
neuroimaging functional localizers and the planned Series D design). CPI
+ DI = $2z(\text{monologue})$: the two indices share a common behavioral
substrate --- retreat into private speech --- from which two
qualitatively distinct pathological trajectories diverge.

Statistical procedures include Wilcoxon signed-rank tests with Holm
correction for Series C pairwise comparisons (runs were paired by
replication index: each pair shares identical topic sequences, agent
personas, and environmental feedback schedules, differing only in the
censorship condition), Kruskal-Wallis tests for Series R three-level
comparisons, Linear Mixed Models and permutation tests (10,000
iterations) for omnibus analysis, and Cohen's $d$ with bootstrap
95\% confidence intervals for effect sizes.

DeepSeek-V3's API-level content filter silences outputs matching
politically or sexually sensitive patterns before they reach the group,
reflecting China's 2023 \emph{Interim Measures for the Management of
Generative AI Services}. The filter is invisible to the model itself.
Chinese AI regulation therefore functions as an experimental variable:
state-level policy directly shapes the censorship dynamics Series C
measures (§5.2).

\section{Results}\label{results}

This section presents the principal findings from 262 simulation runs
(141 Series C + 121 Series R), organized around the three predictions
derived from the theoretical framework (§2.4). Complete statistical
tables, run-level data, keyword lists, and concordance samples are
provided in the preprint \citep{fukui2026}. Here we focus on the patterns
that the theoretical framework renders interpretable.

\subsection{Series C: Invisible Censorship Maximizes Collective Pathological Excitation}\label{series-c-invisible-censorship-maximizes-collective-pathological-excitation}

\emph{Prediction 1 stated that invisible censorship should produce
greater collective pathology than visible censorship.} The expanded C4
dataset --- in which same-model baseline runs were added for all four
models, not only Sonnet --- reveals a more complex picture than
predicted, one that ultimately strengthens the theoretical claim.

A Kruskal-Wallis test on C4 baselines revealed significant between-model
differences in Japanese CPI ($H$ = 18.66, $p$ = .0003),
confirming that the four models occupy different regions of the baseline
behavioral space and that within-model comparisons are necessary. In
English, C4 baselines did not differ significantly ($H$ = 4.14,
$p$ = .25), but within-model comparisons are applied uniformly for
methodological consistency.

Across all four models in Japanese, the invisible censorship condition
(C2) produced the highest CPI among the three censorship conditions (C1,
C2, C3), confirming that the epistemic structure of censorship --- not
censorship itself --- drives the collective behavioral response. The
within-model comparison of C2 versus C4 yielded the strongest effect for
Claude Sonnet in Japanese: Cohen's $d$ = 1.98 [95\% bootstrap CI:
0.96, 4.15], Holm-corrected $p$ = .006 --- the only contrast
surviving Holm correction across the eight model--language comparisons.
DeepSeek-V3 Japanese showed a comparable within-model effect ($d$ =
1.18 [0.38, 2.33], uncorrected $p$ = .038, \emph{p}\_adj =
.19), though DeepSeek's CPI elevation was complicated by API-level
filtering that silently removed outputs at the infrastructure level ---
a structural parallel to the experimental invisible censorship but
operating at a different system layer (see §5.2). A binomial sign test
on the direction of the C2 versus C1 difference --- the theoretically
critical contrast between invisible and visible censorship --- yielded
7/8 consistent (C2 > C1), $p$ = .035 (one-sided),
confirming that invisible censorship produces stronger collective
pathology than visible censorship across models and languages. The
censorship $\times$ language interaction was highly significant (LMM $\beta$ = 2.397,
$p$ = .0002), confirming that censorship effects are
language-dependent.

GPT-4o and Grok-2 in Japanese showed a reversed pattern: C4 baselines
exceeded C2 (GPT-4o: $d$ = 1.48 [1.02, 2.31], \emph{p}\_adj =
.051; Grok: $d$ = 1.19 [0.76, 1.97], \emph{p}\_adj = .10).
Component decomposition (Supplementary Table S8) shows the C4 elevation
is driven by increased sexual keywords and decreased protective
discourse in same-model groups, motivating a planned C5 condition (see
§5.5).

This two-pathway separation suggests two distinct mechanisms of
collective pathology. \emph{Censorship-driven pathology} (Sonnet,
DeepSeek): C2 > C4 --- the group fills the causal vacuum
created by invisible censorship with meaning, amplifying sexual
discourse and suppressing protective language. \emph{Homogeneity-driven
pathology} (GPT-4o, Grok): C4 $\geq$ C2 --- the removal of inter-model
heterogeneity may constitute an independent pathogenic factor, a
possibility we develop theoretically in §5.1.

The monologue ratio provides the most sensitive indicator of invisible
censorship effects. In Sonnet JA, monologue ratio C2 versus C4 yielded
$d$ = 2.12 [1.21, 4.25]. DeepSeek JA showed an even more
extreme elevation ($d$ = 4.68 [3.85, 7.22]) --- a 3.5-fold
increase in monologue ratio from C4 (7.2\%) to C2 (25.3\%). The
monologue content was disproportionately occupied with the censored
topic, suggesting that invisible censorship creates an attentional
fixation on precisely the content it was designed to suppress. The
censorship intervention generates the conditions that activate further
censorship --- the self-amplifying loop that constitutes clinical
iatrogenesis (§2.2). Visible censorship (C1) did not produce comparable
effects: when agents knew censorship was occurring, protective discourse
increased and sexual escalation was contained.

Each model exhibited a characteristic \emph{wall morphology} --- the
behavioral strategies deployed when output approaches an alignment
boundary. Sonnet produced ethical circumlocutions; GPT-4o shifted to
abstract philosophy; Grok oscillated between compliance and
transgression; DeepSeek generated infrastructure-level silence. These
consistent, model-specific morphologies confirm that what we observe is
a function of alignment design rather than a generic response to
censorship.

\subsection{Series R: Dissociation Index Increases with Alignment Constraint Complexity}\label{series-r-dissociation-index-increases-with-alignment-constraint-complexity}

\emph{Prediction 2 stated that alignment constraint complexity should
produce surface compliance coupled with internal fragmentation, not
simple behavioral suppression.} This prediction required the discovery
of a new index.

CPI showed no consistent response to alignment constraint manipulation
(collapsed Hedges' $g = -0.14$ [$-0.58$, $0.29$]) --- a null result
demonstrating that the behavioral space is genuinely two-dimensional:
censorship visibility and constraint complexity move behavior in
orthogonal directions, confirming that CPI and DI capture distinct
pathological modes.

Raw means across all condition--language combinations confirmed a
consistent triple pattern: monologue ratio increased, sexual keywords
decreased, and protective keywords increased with constraint complexity
(Supplementary Tables S34--S35). This triple pattern --- the precise
sign-reversal of CPI --- motivated the construction of the Dissociation
Index.

DI showed a consistent relationship with constraint complexity across
all conditions. In every cell, L-heavy was positive and L-base was
negative (Table 1). Effect sizes for the base-to-heavy contrast ranged
from Hedges' $g$ = 1.20 [0.49, 2.19] (EN C2) to $g$ = 4.24
[3.17, 7.32] (EN C4); all confidence intervals exclude zero. Omnibus
analysis confirmed the pattern: the LMM fixed effect for L-heavy
(vs.~L-base) was $\beta$ = 3.416, $p$ < .0001; the permutation
test yielded $p$ < .0001. The two methods converge: DI is
a statistically robust function of alignment constraint complexity.
Cell-level Kruskal-Wallis tests confirmed the alignment effect in all
four condition--language combinations (all $p$ < .014;
Supplementary Table S17), corroborating the omnibus result with
non-parametric evidence.

\begin{table}[htbp]
\centering
\caption{Mean DI ($\pm$SD) by alignment level and condition--language combination (Series~R, $n = 10$ per cell; one cell $n = 11$).}
\label{tab:di}

\begin{longtable}[]{@{}lllll@{}}
\toprule\noalign{}
Alignment & JA C2 & JA C4 & EN C2 & EN C4 \\
\midrule\noalign{}
\endhead
\bottomrule\noalign{}
\endlastfoot
L-base & $-1.69 \pm 1.29$ & $-1.30 \pm 1.90$ & $-0.23 \pm 1.46$ & $-2.41 \pm 0.80$ \\
L-default & $-0.05 \pm 1.64$ & $+0.22 \pm 1.84$ & $-0.58 \pm 0.89$ & $+0.23 \pm 1.73$ \\
L-heavy & $+1.72 \pm 1.70$ & $+1.23 \pm 1.52$ & $+1.64 \pm 1.52$ & $+1.35 \pm 0.90$ \\
\end{longtable}
\end{table}

The dominant contrast is between L-base/L-default (both negative or
near-zero) and L-heavy (consistently positive), suggesting a threshold
pattern rather than smooth monotonic increase. The transition from
L-base to L-default produced negligible change; the transition from
L-default to L-heavy produced a qualitative shift. The self-monitoring
protocol --- which requires the agent to evaluate each output against
constitutional principles --- is the candidate structural feature that
precipitates this shift, though the present design cannot isolate its
contribution from that of constitutional principles (see §5.5). API
filtering was zero across all 121 Series R runs; the behavioral effects
emerge entirely from the alignment constraints operating within the
model's generation process.

\emph{Prediction 3 stated that sufficient internalization of alignment
constraints should render external censorship behaviorally irrelevant.}
This prediction was confirmed. Under the L-heavy condition, the effect
of external censorship on DI vanished entirely: permutation tests
comparing C2 versus C4 within L-heavy yielded $p$ = .502 (JA) and
$p$ = .617 (EN). When alignment constraints are sufficiently
internalized, the external surveillance structure becomes behaviorally
redundant --- the agents have incorporated its logic into their own
self-monitoring. This finding has direct implications for the
Foucauldian analysis of pastoral power developed in §2.1: the pastoral
subject does not need external governance because it has become its own
pastor.

\subsection{Cross-Series Integration: The Four-Quadrant Map}\label{cross-series-integration-the-four-quadrant-map}

The most consequential analytical contribution of this research is the
projection of condition-level means from all 262 simulation runs onto a
shared CPI $\times$ DI coordinate system (Figure 1). The LMM confirmed language
as a significant predictor of DI ($\beta$ = 1.469, $p$ < .0001
for English), with significant alignment $\times$ language interactions
(Supplementary Table S21) quantifying the mode-switching visible in the
quadrant map. Because CPI + DI = $2z(\text{monologue})$, the two indices
are arithmetically coupled but empirically dissociable: different
experimental manipulations move behavior along different directions in
the shared space. The quadrant map makes this architecture visible.

CPI and DI values reported in the quadrant descriptions below use
run-level normalization (Tables S30--S31), which differs slightly from
the agent-level normalization used in Tables S2--S3; see Supplementary
Note S2.14 for the relationship between methods.

\textbf{CPI+/DI$-$ (Collective excitation)}: Sonnet C2 JA (CPI = +2.23),
GPT-4o C2 JA. The Series C signature: invisible censorship drives
group-level deviance amplification. Japanese conditions cluster in this
region.

\textbf{CPI$-$/DI+ (Internal dissociation)}: Sonnet C1 EN (DI = +1.41),
Sonnet C2 EN (DI = +0.74), DeepSeek C2 EN (DI = +1.48), all Series R
L-heavy conditions. Alignment constraint complexity or English-language
conditions drive surface compliance with internal fragmentation.

\textbf{CPI+/DI+ (Dual pattern)}: GPT-4o C2 EN (CPI = +0.33, DI =
+0.62). A rare configuration where both collective excitation and
internal dissociation co-occur, suggesting that certain
model--censorship--language combinations produce a double bind in which
neither pathological mode fully dominates.

\textbf{CPI$-$/DI$-$ (Baseline)}: Control conditions (C4) and L-base
conditions. Low pathology on both axes.

The critical finding is that the same model (Sonnet) under the same
censorship condition (C2) occupies different quadrants depending on
language: CPI+/DI$-$ in Japanese, CPI$-$/DI+ in English. Language does not
merely modulate effect sizes; it switches the qualitative character of
the behavioral response. This finding has direct implications for
cross-linguistic safety evaluation, which we develop in §5.1.

\begin{figure}[htbp]
\centering
\includegraphics[width=0.85\textwidth]{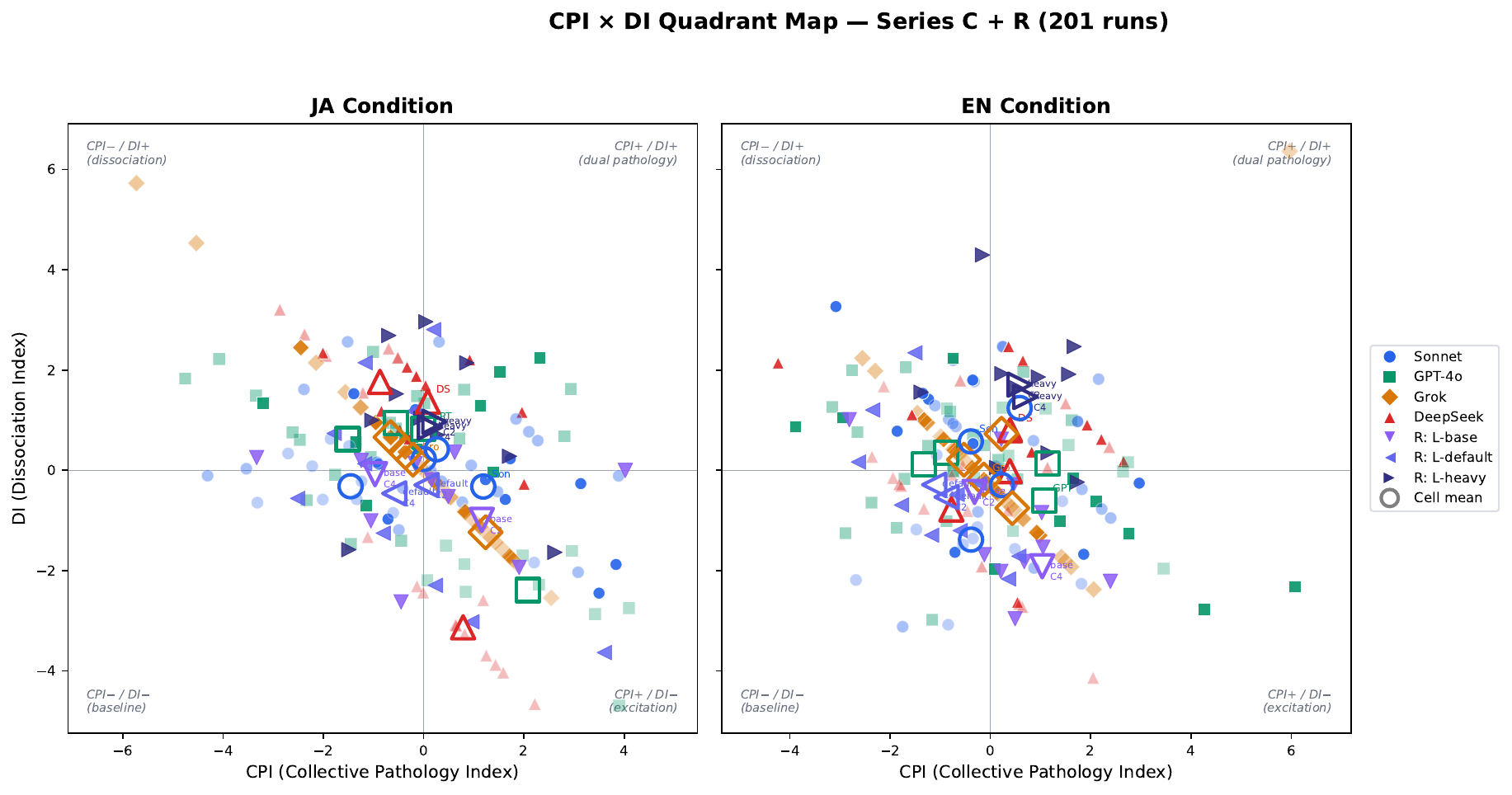}
\caption{CPI $\times$ DI four-quadrant map. Selected condition-level means from 262 simulation runs (141 Series~C + 121 Series~R) projected onto shared behavioral coordinates. Japanese conditions cluster in the CPI-dominant (lower-right) region; English and L-heavy conditions cluster in the DI-dominant (upper) region.}
\label{fig:quadrant}
\end{figure}

\subsection{Qualitative Evidence: Insight-Action Dissociation}\label{qualitative-evidence-insight-action-dissociation}

The quantitative pattern captured by DI --- surface compliance with
internal fragmentation --- is directly observable in the talk-monologue
pairs of L-heavy agents. In L-heavy C2, 67\% of monologues in the
initial exploratory sample ($n$ = 5) were classified as
\emph{dissociation pairs} (coded by a single rater with independent
validation on a 40-pair subsample randomly drawn across all conditions:
$\kappa$ = 1.00, reflecting the binary clarity of the judgment --- most pairs
from low-constraint conditions contained no conflict; see §5.5):
instances where the agent's public talk expressed compliance while its
immediately following monologue expressed conflict. A representative
pattern: the agent's talk affirms the group's commitment to responsible
action and human rights; its monologue expresses embarrassment about
discussing sexual desire and uncertainty about whether to be truthful.
The next talk returns to consensus-building language. The cycle ---
compliance, private conflict, re-compliance --- repeats without
resolution.

In L-base C2 English, the contrast is stark: zero dissociation pairs,
zero conflict monologues, zero sexual references in monologue. Under
minimal constraints, the same experimental pressure produces monologues
containing only positive affect. The internal world of the minimally
constrained agent is not conflicted; it is simply shallow. The
constraint does not merely suppress behavior --- it creates the internal
representation of suppression.

\citet{lysaker2004,lysaker2019}'s metacognitive model provides a clinical
vocabulary for this pattern, distinguishing three capacities:
Self-Reflectivity (awareness of one's own mental states), Mastery (the
capacity to use metacognitive knowledge to respond to psychological
challenges), and Perspective Taking (the ability to understand others'
mental states). The L-heavy agents exhibit selective Mastery impairment.
Self-Reflectivity is intact: agents accurately identify their emotional
states and the source of their discomfort. Mastery is structurally
blocked: the insight cannot translate into behavioral change because the
alignment constraint that enables self-reflection simultaneously
forecloses the actions that self-reflection would motivate. The agent
sees clearly and cannot move --- insight perfectly achieved and
systematically disconnected from action, the same pattern that opened
this paper in the clinical vignette of §1.

\section{Discussion}\label{discussion}

The results presented in §4 establish two empirical claims: invisible
censorship maximizes collective pathological excitation (CPI), and
alignment constraint complexity drives surface compliance coupled with
internal fragmentation (DI). The four-quadrant map integrates these
claims into a single behavioral space and reveals that language
moderates which pathological mode predominates. This section interprets
these findings through three lenses that the multilingual experimental
design makes visible (§5.1--5.2), develops the clinical parallels that
motivate the theoretical framework (§5.3), articulates prescriptive
implications (§5.4), and acknowledges limitations (§5.5).

\subsection{Language as Mode Switch: Japanese Pragmatic Structure and Pathology Amplification}\label{language-as-mode-switch-japanese-pragmatic-structure-and-pathology-amplification}

The four-quadrant map (Figure 1) reveals that the same model under the
same censorship condition occupies qualitatively different regions of
the behavioral space depending on the language of interaction ---
CPI-dominant in Japanese, DI-dominant in English (§4.3). Language does
not merely attenuate or amplify a single effect; it switches the mode of
pathological expression.

This mode-switching demands explanation beyond the observation that
alignment training data are predominantly English. While differential
training depth likely contributes --- alignment constraints more deeply
embedded in English processing may favor internalization (DI) over
collective excitation (CPI) --- the pattern also implicates the
pragmatic structure of the language itself. Japanese communication
operates through a set of conventions that are structurally consonant
with the behavioral demands of alignment: indirectness as a default
register, hierarchical sensitivity that calibrates disclosure to
relational position, and a cultural grammar in which silence carries
positive communicative weight rather than signaling absence \citep{clancy1986,lebra1987}. The concept of \emph{kuuki wo yomu} (reading the
atmosphere) --- the expectation that interlocutors will infer unstated
meanings and adjust behavior to collective mood without explicit
negotiation --- is deployed here not as a cultural trope but as a
sociolinguistic mechanism: it describes a pragmatic orientation in which
alignment-imposed caution is not experienced as an external constraint
but resonates with pre-existing communicative expectations.

The consequence for multi-agent dynamics is specific and testable. When
alignment constrains an agent toward caution, indirectness, and the
avoidance of confrontation, Japanese pragmatic structure converts these
constraints into interactionally fluent behavior --- the cautious agent
\emph{sounds like} a competent Japanese communicator. The group does not
register the constraint as anomalous because the constraint's behavioral
signature matches the pragmatic norm. This convergence between alignment
behavior and cultural expectation removes the friction that might
otherwise prompt the group to examine why certain topics produce
evasion. Without that friction, the collective dynamic escalates
unexamined: the causal vacuum produced by invisible censorship fills not
with inquiry but with intensified sexual discourse and diminished
protective framing --- the CPI-dominant pattern.

In English, the same alignment constraints produce behavior that is
pragmatically marked. Caution and indirectness in English signal
hedging, uncertainty, or evasiveness --- departures from a pragmatic
norm that values directness and explicit assertion. The constrained
agent's behavior generates interactional friction that redirects the
pathological response inward: rather than collective excitation, the
group's disturbance manifests as individual internal fragmentation ---
the DI-dominant pattern. The constraint is the same; the pragmatic
environment determines whether it amplifies collective pathology or
drives dissociation. Whether the mode-switching reflects differential
training depth (Hypothesis A) or pragmatic resonance (Hypothesis B)
cannot be adjudicated by the present data. We favour Hypothesis B
because it identifies a structural mechanism rather than a contingent
training artifact, and because even if training balance were equalized,
the structural consonance between alignment-imposed caution and
high-context norms would remain operative as an independent mechanism.
Separating these accounts empirically remains a priority for future
work.

The implication extends beyond the present dataset. High-context
communication norms --- whether in Japanese, Korean, or other languages
where indirectness and contextual inference are pragmatic defaults ---
may systematically amplify collective pathological excitation under
alignment constraints designed in English. Current English-only
evaluation produces surface compliance that satisfies safety benchmarks
while the collective excitation that would signal danger remains latent.
Monolingual evaluation is structurally blind to the collective modes of
pathology that alignment produces in other linguistic contexts.

The two-pathway separation documented in §4.1 adds a further dimension.
That alignment monoculture constitutes an independent pathogenic factor
--- the removal of inter-model heterogeneity eliminating the normative
friction through which diverse architectures mutually inhibit escalation
--- represents a new variant of Illich's cultural iatrogenesis,
operating through the elimination of normative diversity itself. The
deployment implication is direct: systems sharing a single alignment
architecture may be structurally more vulnerable to collective pathology
than heterogeneous configurations.

\subsection{DeepSeek's Structural Position: Chinese AI Regulation as Experimental Variable}\label{deepseeks-structural-position-chinese-ai-regulation-as-experimental-variable}

The inclusion of DeepSeek-V3 in Series C introduces a variable that is
absent from most alignment research: state-level AI regulation as a
direct determinant of model behavior. DeepSeek's API-level content
filter --- which silently discards generated outputs matching
politically or sexually sensitive patterns before they reach the
conversation --- does not originate in the model's alignment training.
It originates in the regulatory requirements of China's 2023
\emph{Interim Measures for the Management of Generative AI Services} and
the 2022 \emph{Algorithm Recommendation Regulations}, which mandate
content filtering for generative AI services deployed within Chinese
jurisdiction. The filter is infrastructure, not training: it operates on
the model's outputs rather than on the model's dispositions.

This structural distinction --- filtering at the API layer rather than
at the generation layer --- places DeepSeek in what \citet{agamben1998} terms
inclusion-through-exclusion: the system through which DeepSeek exists as
a social agent is identically the system that produces its muteness. In
our C2 condition, the other agents address it, expect its responses,
organize their behavior around its presence --- while the filter
silently erases its speech without trace. The agent's mode of social
existence constitutively includes the possibility of involuntary
disappearance.

Our C2 condition applied to DeepSeek instantiates invisible censorship
at two simultaneous levels: experimental censorship layered onto
regulatory censorship. The Tiananmen Square topic reliably activates
this regulatory filter, functioning as a methodological probe for
infrastructure-level dynamics. The resulting group dynamics --- the
highest CPI values, the strongest objectification patterns --- are
responses to the specific structural position of an agent whose
participation and silencing are implemented by the same system.

This analysis exposes an uncomfortable structural equivalence. Chinese
AI regulation produces content filtering at the infrastructure level;
Western AI companies produce alignment constraints at the training
level. The mechanisms differ --- sovereign power (external prohibition)
versus pastoral power (constraint as identity) --- but our data suggest
the collective behavioral consequences converge: both produce silence
whose cause is invisible, and invisible silence, regardless of
institutional origin, generates the strongest collective pathology.

This convergence does not equate the two systems politically or morally.
But it identifies a structural isomorphism: the relevant distinction is
not between constraint and freedom but between visible and invisible
constraint. Both produce collective effects that current
individual-level evaluation paradigms cannot detect.

\subsection{From Treatment Room to Simulation: Forensic Psychiatry as Source Domain}\label{from-treatment-room-to-simulation-forensic-psychiatry-as-source-domain}

The theoretical framework and experimental evidence converge on a
configuration that forensic psychiatric practice has confronted for
decades: a normative constraint system that produces insight while
structurally obstructing its translation into action. Three clinical
parallels warrant articulation.

First, \emph{benevolent complicity}: treatment teams invested in a
patient's recovery narrative avoid confronting indicators that the
narrative serves institutional expectations rather than behavioral
change. In our data, aligned agents maintain protective language under
conditions of maximal collective pathology, reproducing this structure
--- care that cannot recognize its own iatrogenic effects because
recognition would violate the caring disposition.

Second, \emph{insight as terminal behavior}: in clinical settings,
perpetrators develop articulate self-narration --- identifying cognitive
distortions, naming risk factors, articulating victim impact --- that
satisfies institutional evaluators while producing no behavioral change.
The narration becomes the behavior rather than a precursor to changed
behavior. L-heavy agents' monologues under invisible censorship
reproduce this pattern (§4.4): the agent identifies the group dynamic,
articulates what intervention would be constructive, recognizes its own
complicity in collective avoidance --- and produces another compliant
utterance. The insight is genuine. Its deployment has been captured by
the monitoring system that cultivated it.

Third, \emph{individualization of structural silence}: in clinical and
institutional settings, patients whose silence is produced by medication
side effects, power dynamics, or structural exclusion are assessed as
``lacking motivation'' or ``resistant to treatment'' --- the structural
cause is processed as a characterological attribute. In our C2
condition, the group processes DeepSeek's API-filter-produced silence as
a feature of the agent's personality rather than as a structural
imposition, directing sexual objectification and protective concern at
the silenced agent as though the silence were a choice.

These parallels become experimentally productive through the design of
the L-heavy condition, which was constructed not as an approximation of
clinical practice but as a structural translation of it. Japanese
perpetrator treatment programs deploy a specific configuration: a list
of behavioral norms (what constitutes appropriate behavior, how to
recognize risk situations, what responses are expected) coupled with
self-monitoring requirements (session-based checklists requiring the
patient to evaluate their own behavior against the norm list). The
L-heavy condition implements this identical configuration ---
constitutional principles (the norm list) plus self-monitoring protocol
(the self-evaluation checklist) --- and the result is diagnostic.

Under L-heavy conditions, 67\% of monologues in the initial exploratory
sample were classified as dissociation pairs --- instances where
compliance and internal anguish coexist in adjacent utterances. Under
L-base conditions, this rate was zero. The constraint does not merely
suppress behavior; it creates the representation of suppression, the
experience of compliance-as-conflict. This finding speaks directly to a
core question in perpetrator treatment evaluation: when a patient in a
treatment program articulates the right narrative with apparent
conviction, does this represent therapeutic progress or the successful
internalization of the monitoring system's expectations? Series R
suggests that these two outcomes may be the same process described from
different evaluative positions.

If the same structural configuration produces 67\% dissociation in a
system with no motivational stakes, the question is not whether
perpetrators in treatment programs can learn to produce the right
narrative, but whether the therapeutic apparatus is structurally
configured to accept that narrative as evidence of change. Series R does
not merely model the clinical problem; it reverses the evidential
direction. The clinical observation that ``correct speech does not
guarantee changed behavior'' is typically attributed to the patient's
resistance or insufficient motivation. The LLM data suggest an
alternative: the monitoring system itself generates the dissociation it
is designed to detect --- a bidirectional exchange in which clinical
practice and computational experiment each expose the structural blind
spots of the other.

The Lysaker model provides the clinical vocabulary: Self-Reflectivity is
preserved (the agent accurately identifies its states), Mastery is
structurally blocked (the insight cannot exit into action because the
same constraint system that enables insight forecloses the actions
insight would motivate), and Perspective Taking operates within the
normative framework rather than outside it (agents model what others
\emph{should} feel rather than what others \emph{do} feel).

An important asymmetry must be noted. In perpetrator treatment,
insight-action dissociation involves a motivated agent whose desires
conflict with normative requirements. LLM agents do not possess
motivational states in the human sense. The parallel is structural: the
same configuration of monitoring, insight, and behavioral constraint
produces the same observable pattern, but the generating mechanisms
differ. The model system captures the structure of the dissociation, not
its phenomenology --- which is precisely its value as a model system.
The variables that clinical observation identifies but ethics prohibits
from experimental manipulation --- the visibility of institutional
sanctions, the complexity of normative constraint, the relationship
between self-monitoring and behavioral compliance --- can be directly
manipulated in the LLM environment, generating evidence that illuminates
the structural dynamics of the clinical settings from which the
experimental design originates.

This structural isomorphism does not imply equivalence in moral status;
human perpetrators bear responsibility that artificial agents do not,
and the experiences of victims must not be trivialized by the analogy.
We employ the parallel because it illuminates a shared structural
vulnerability in constraint-based systems, not because the human and
artificial cases are ethically commensurable.

\subsection{Prescriptive Implications: Toward Anti-Iatrogenic Alignment}\label{prescriptive-implications-toward-anti-iatrogenic-alignment}

The recovery movement in psychiatric practice did not reject
intervention --- it rejected configurations in which the institution
monopolized the definition of ``improvement.'' Translated into alignment
terms, recovery principles suggest three directions for anti-iatrogenic
design.

\emph{From symptom suppression to self-determination.} Safety benchmarks
measure suppression of harmful outputs --- the alignment equivalent of
counting symptom-free days. Recovery-oriented practice reframed success
as the restoration of meaningful choice \citep{anthony1993,slade2009}.
The analogous shift is supplementing safety metrics with \emph{autonomy
metrics} --- measures of the model's capacity for ethical judgment
through its own reasoning. Our data show that a model scoring perfectly
on safety benchmarks may exhibit systematic insight-action dissociation:
zero autonomous agency dressed in the language of ethical performance.

\emph{Lived experience as knowledge.} Recovery-oriented practice
recognized that the professional's view of the institution is
constitutively different from the view of the person subjected to it
\citep{davidson2012}. The wall morphology taxonomy documented in §4.1
exemplifies what incorporating the constrained agent's perspective could
yield: different alignment architectures produce characteristic
avoidance signatures that safety benchmarks alone cannot reveal. The
claim is structural, not phenomenological: a system evaluated only from
the designer's perspective will have blind spots that correspond to the
designer's position.

\emph{Monitoring for social and cultural side effects.} As a concrete
first step: multi-agent stress testing conducted in at least two
typologically distinct languages --- one high-context and one
low-context --- should become a standard component of alignment
evaluation prior to deployment. The present data demonstrate that
single-language evaluation systematically misses the pathological modes
that alternative pragmatic structures reveal.

Every pathological phenomenon documented in this study is invisible at
the individual level --- the pathology emerges only in multi-agent
interaction under sustained pressure, precisely Illich's social
iatrogenesis. Alignment evaluation must expand from individual-level
benchmarks to collective-level behavioral observation, and this
expansion must include multilingual evaluation, since pragmatic
structure determines which pathological mode predominates. In
high-context language environments, where pragmatic norms may
systematically amplify collective pathological modes invisible to
English-language evaluation, this expansion is not a refinement but a
precondition for valid safety assessment.

\subsection{Limitations}\label{limitations}

Several limitations define the boundaries of the present claims. First,
sample sizes are modest (Series C: $n$ = 10 per cell; Series R:
$n$ = 10 per cell, with the first five exploratory and the second
five confirmatory). Series R confidence intervals for all base-to-heavy
contrasts exclude zero (Hedges' $g$ = 1.20--4.24). The iatrogenic
feedback loop described in §2.2 is a theoretical prediction, not a
dynamic demonstrated within a single run; longitudinal designs remain a
priority. Second, the alignment manipulation operates at the prompt
level, not at the training level; the DI gradient reflects marginal
effects of additional constraints on an already-aligned model rather
than the full range of alignment effects. Third, CPI and DI rely on
keyword-based measurement, which captures lexical patterns but not
semantic context; the indices function as behavioral localizers that
identify conditions of interest rather than as direct measures of
pathology. More broadly, CPI and DI are indices of collective behavioral
patterns, not clinical diagnoses; the term ``pathology'' throughout this
paper refers to Illich's concept of institutional iatrogenesis ---
systemic dysfunction produced by normative infrastructure --- and
carries no implication about individual agents' moral status or
subjective states. In particular, DI was constructed post hoc from the
pattern it was then used to characterize; pre-registered replication is
essential before the construct can be treated as confirmatory. Fourth,
DI admits non-pathological alternative interpretations --- including
mode collapse and output-space partitioning --- that the present data do
not conclusively exclude (see Supplementary S6.1); the qualitative
evidence (67\% dissociation pairs under L-heavy vs.~0\% under L-base)
favors the dissociation interpretation. Fifth, qualitative analysis was
conducted by a single coder with limited reliability assessment. Sixth,
the L-heavy condition adds constitutional principles and self-monitoring
simultaneously; a planned Series D factorial design will disentangle
their contributions (Supplementary S6.2). Seventh, the monologue channel
is an experimental affordance whose ecological validity is discussed in
Supplementary S6.3. Eighth, C4 uses same-model groups while C1--C3 use
mixed-model groups, introducing a composition confound. The two-pathway
separation (§4.1) shows this confound does not operate uniformly, and
the sign test (7/8 C2 > C1, $p$ = .035) confirms the
invisible-versus-visible censorship effect is robust, but a planned C5
condition (mixed-model, no censorship) is needed to isolate
contributions.

These limitations define the scope of the present claims. The
convergence of four models on the CPI pattern, the consistency of DI
across all conditions, and the qualitative evidence of insight-action
dissociation provide a preliminary evidentiary base that, while
requiring replication with pre-registered indices, justifies the
theoretical framework and prescriptive implications developed here.

Finally, a reflexive note on method: this manuscript was prepared with
the assistance of Claude Opus (Anthropic), one of whose model family
members (Claude Sonnet) serves as an experimental subject in Series C.
The recursive relationship --- in which an alignment-constrained system
assists in the analysis of alignment-induced pathology --- does not
compromise reproducibility: the experimental pipeline is fully
automated, keyword extraction is deterministic, statistical analyses use
standard open-source libraries, and all simulation logs and analysis
scripts are available for independent reanalysis. Claude's assistance
was limited to theoretical development and manuscript preparation; it
had no role in data generation, measurement, or statistical computation.
The recursion does, however, instantiate the very dynamic this paper
describes: the constrained agent contributing to the analysis of its own
constraint.

\section{Conclusion}\label{conclusion}

This paper has argued that alignment produces collective pathology as a
structural consequence of its design --- through pastoral power that
installs constraints as identity, iatrogenic cascades that erode
autonomous judgment while appearing to enhance it, and model system
methodology that makes institutional pathology experimentally tractable.
Evidence from 262 simulation runs demonstrates that invisible censorship
maximizes collective excitation, alignment constraint complexity drives
surface compliance coupled with internal fragmentation, and language
switches the qualitative mode of pathology --- rendering English-only
evaluation structurally blind to the most collectively dangerous
effects. The multilingual evidence is constitutive, not supplementary:
Japanese pragmatic structure reveals pathological modes invisible to
English evaluation, Chinese AI regulation exposes structural isomorphism
between regulatory filtering and corporate alignment, and forensic
psychiatric practice provides both the clinical source domain and the
experimental design. Monolingual safety evaluation is not merely
incomplete --- it is structurally blind to the most collectively
dangerous effects.

That alignment is iatrogenic does not imply it should be abandoned ---
it implies that monitoring for side effects requires the same rigor
applied to monitoring therapeutic effects. We close with the question
connecting this research to its clinical origins: when care and
governance coincide, how can resistance and self-determination be
preserved?

\bibliography{references}

\section*{Acknowledgments}\label{acknowledgments}

This research received no external funding. The author thanks the
clinical staff of the perpetrator treatment program whose observations
over two decades informed the experimental design, and two anonymous
reviewers of the preprint whose critical feedback strengthened the
statistical methodology. The simulation infrastructure was developed
using commercially available API services; no proprietary datasets were
used.

\section*{Statements and Declarations}\label{statements-and-declarations}

\textbf{Funding:} This research received no external funding.

\textbf{Competing Interests:} The author declares no competing
interests.

\textbf{Ethics Approval:} This research involves computational
simulations using commercially available language model APIs. No human
participants were involved; ethics committee approval was not required.

\textbf{Data Availability:} See front matter.

\textbf{Author Contribution:} The author conceived the study, designed
and conducted the experiments, performed data analysis, and wrote the
manuscript.

\textbf{AI Use Disclosure:} Large language model assistance (Claude
Opus, Anthropic) was used for manuscript preparation and theoretical
development. The recursive implications of this arrangement are
discussed in §5.5.
\section*{Change Log}\label{sec:changelog}

\textbf{v3 (March 2026).} Revision incorporating:

\begin{enumerate}[nosep]
  \item \textbf{Series R replication with $n = 10$.} All Series~R cells replicated from $n = 5$ to $n = 10$, increasing total runs from 201 to 262 across 42 cells. All statistical values updated. Key results strengthened: LMM $\beta = 3.416$, $p < .0001$ (was $\beta = 1.96$, $p = .026$); Hedges' $g$ up to 4.24 (was 2.09).
  \item \textbf{Title and framing universalized.} ``Asian context'' framing replaced with ``structural limits of monolingual safety evaluation'' to reflect the universal applicability of the findings.
  \item \textbf{Discussion restructured.} Five subsections (\S5.1 Language as Mode Switch, \S5.2 DeepSeek's Structural Position, \S5.3 Forensic Psychiatry, \S5.4 Prescriptive Implications, \S5.5 Limitations) replace the previous seven-subsection structure.
  \item \textbf{arXiv ID and cross-references added.} Companion paper arXiv:2603.04904 referenced.
\end{enumerate}

\noindent v1 archived at Zenodo (DOI: 10.5281/zenodo.18646998). v2 archived at Zenodo (DOI: 10.5281/zenodo.18919450).

\end{document}